\documentclass[aps,twocolumn,pra,tightenlines,floatfix,showpacs]{revtex4}
\usepackage[dvips]{graphicx}
\usepackage[english]{babel}
\usepackage{amsmath}
\usepackage{amssymb}
\usepackage{times}

\newcommand{\bs}{\begin{split}}
\newcommand{\es}{\end{split}}
\newcommand{\be}{\begin{equation}}
\newcommand{\ee}{\end{equation}}
\newcommand{\ba}{\begin{eqnarray}}
\newcommand{\ea}{\end{eqnarray}}

\def\ket#1{|#1\rangle}

\begin{document}

\title{Radio frequency spectroscopy and the pairing gap in trapped
  Fermi gases }

\author{Yan He, Qijin Chen, and K.  Levin}

\affiliation{ James Franck Institute and Department of Physics,
  University of Chicago, Chicago, Illinois 60637}

\date{\today}

\begin{abstract}

  We present a theoretical interpretation of radio-frequency (RF)
  pairing gap experiments in trapped atomic Fermi gases, over the
  entire range of the BCS-BEC crossover, for temperatures above and
  below $T_c$.  Our calculated RF excitation spectra, as well as the
  density profiles on which they are based, are in semi-quantitative
  agreement with experiment.  We provide a detailed analysis of the
  physical origin of the two different peak features seen in RF
  spectra, one associated with nearly free atoms at the edge of the
  trap, and the other with (quasi-)bound fermion pairs.

\end{abstract}

\pacs{03.75.Hh, 03.75.Ss, 74.20.-z \hfill \textsf{\textbf{cond-mat/0504394}}}

\maketitle

%% -----------------------------------

A substantial body of experimental evidence for superfluidity in trapped
fermionic gases \cite{Jin4,Ketterle3a,Thomas2a,Grimm3a} has focused
attention on an important generalization of BCS theory associated with
arbitrarily tunable interaction strengths; this is called
``BCS--Bose-Einstein condensation (BEC) crossover theory"
\cite{ourreview}. This tunability is accomplished via magnetic field
sensitive Feshbach resonances.  At weak interaction strength
conventional BCS theory applies so that pairs form and condense at the
same temperature, $T_c$, whereas as the attraction becomes strong, pairs
form at one temperature ($T^*$) and Bose condense at another ($T_c <
T^*$).  The intermediate or unitary scattering regime, (where the
fermionic two-body $s$-wave scattering length $a$ is large), is of
greatest interest because it represents a novel form of fermionic
superfluidity. In contrast to the BEC case, there is an underlying Fermi
surface (in the sense that the fermions have positive chemical potential
$\mu$), but ``pre-formed pairs'' are already present at the onset of
their condensation.

The difficulty of obtaining phase sensitive probes and the general
interest in this novel superfluidity make experiments which probe the
fermionic excitation gap extremely important. While in the weak
coupling BCS limit the gap onset appears at $T_c$, in the unitary
regime this gap (or ``pseudogap") appears at a high temperature $T^*$
and directly reflects the formation of (quasi-)bound fermion pairs
\cite{JS2a,ourreview,Grimm4a,Jin5}.  For the trapped Fermi gases, one
has to devise an entire new class of experiments to measure this
pairing gap; traditional experiments, such as superconductor-normal
metal (SN) tunneling are neither feasible nor appropriate.  The first
such experiment was based on radio frequency (RF) spectroscopy
\cite{Grimm4a}; this followed an earlier proposal by Kinnunen
\textit{et al} \cite{Torma,Torma1}, who also presented an
interpretation of recent data in $^6$Li \cite{Torma2} in the
\emph{unitary} regime.  However, some issues have been raised about
their interpretation in the literature \cite{Griffin5}. Moreover, the
spectra in the BEC and BCS regimes also need to be addressed.

It is the purpose of the present paper to present a more systematic
analysis of RF pairing gap experiments for the entire experimentally
accessible crossover regime from BCS to BEC, as well as address recent
concerns \cite{Griffin5}.  Our studies address the two peaks in the
spectra observed experimentally at all temperatures, and clarify in
detail their physical origin.  Essential to the present approach is
that our calculations are based on trap profiles \cite{JS5} and
related thermodynamics \cite{ThermoScience} which are in quantitative
agreement with experiment at unitarity \cite{Thomas,ThermoScience}
where there is a good calibration.

In the RF experiments \cite{Grimm4a}, one focuses on three different
atomic hyperfine states of the $^6$Li atom.  The two lowest states,
$\ket1$ and $\ket2$, participate in the superfluid pairing.
%, and are described by the Hamiltonian $H-\mu N$. 
 The higher state, $\ket3$, is
effectively a free atom excitation level; it is unoccupied initially.
An RF laser field, at sufficiently large frequency, will drive atoms
from state $\ket2$ to $\ket3$.

As in Refs. \cite{JS5,ChenThermo} we base our analysis on the conventional
BCS-Leggett ground state \cite{Leggett}, extended
\cite{JS2a,ourreview} to address finite temperature effects and to
include the trap potential.  In this approach, pseudogap effects are
naturally incorporated. We begin with the usual two-channel grand
canonical Hamiltonian $H-\mu N$ \cite{Timmermansa} which describes
states $\ket1$ and $\ket2$, as in Ref.~\cite{ChenThermo}, and solve
for the spatial profiles of relevant physical quantities.  As a result
of the relatively wide Feshbach resonance in $^6$Li, the fraction of
closed-channel molecules is very small for currently accessible
fields.  Therefore, we may neglect their contribution to the RF
current, as was done in Ref.~\cite{Torma}.

The Hamiltonian describing state $\ket3$ is given by $H_3-\mu_3 N_3=
\sum_{\mathbf{k}} (\epsilon_{\mathbf{k}}+\omega_{23}-\mu_3)
c_{3,\mathbf{k}}^{\dagger} c^{}_{3,\mathbf{k}}$, where $\epsilon_{\bf
  k}$ is the atomic kinetic energy, $c^{}_{3,\mathbf{k}}$ is the
annihilation operator for state $\ket3$, $\omega_{23}$ is the energy
splitting between $\ket3$ and $\ket{2}$, and $\mu_3$ is the chemical
potential of $\ket3$.
In addition, there is a transfer matrix element $T_{\bf k,p}$ from
$\ket2$ to $\ket3$ given by
$  H_T  =\sum_{\bf k,p}(T_{\bf k,p}\,c_{3,\bf p}^\dag c_{2,\bf k}^{}
  +h.c.) $
For plane wave states, 
$T_{\bf k,p} = \bar {T }\delta({\bf q}_L+{\bf k}-{\bf p})\delta(\omega_{\bf
  {kp}} - \omega_L)$.
Here $q_L \approx 0$ and $\omega_L$ are the momentum and energy of the
RF laser field, and $\omega_{\bf {kp}}$ is the energy difference between
the initial and final state.  It should be stressed that unlike
conventional SN tunneling, here one requires not only conservation of
energy but also conservation of momentum.

The RF current is defined as $I=\langle \dot{N}_3\rangle=i\langle
[H,N_3]\rangle$. Using standard linear response theory one finds $I=2
\bar{T}^2{\rm Im}[X_{ret}(-\omega_L+\mu_3-\mu)]$. Here the retarded
response function $X_{ret}(\omega) \equiv X(i\omega_n \rightarrow
\omega+i0^+)$, and the linear response kernel $X$ can be expressed in
terms of single particle Green's functions as $X(i\omega_n)= T
\sum_{m,{\bf k}}G_3({\bf k},i\nu_m)G({\bf k+q}_L,i\nu_m+i\omega_n)$, where
$\omega_n$ and $\nu_m$ are even and odd Matsubara frequencies,
respectively. (We use the convention $\hbar=k_B=1$). After Matsubara
summation we obtain
%the result can be written in
%terms of spectral functions as
%
\begin{equation} 
  I=\frac{\bar{T}^2}{2\pi}\int \!\mathrm{d}\nu\sum_{{\bf 
      k},{\bf p}}  A_3({\bf k},\nu)A({\bf p},\nu') 
  \delta({\bf q}_L+{\bf k}-{\bf p})
  \left[f(\nu')-f(\nu)\right] \,,
\end{equation} 
%\ba 
%I&=&\frac{\bar{T}^2}{2\pi}\int_{-\infty}^{\infty}
%\!\mathrm{d}\epsilon\sum_{{\bf 
%    k},{\bf l}}
%A_3({\bf k},\epsilon)\,A({\bf l},\epsilon') \nonumber\\
%&&{}\times\delta({\bf q}_L+{\bf k}-{\bf l})
%\left[f(\epsilon')-f(\epsilon)\right] \,,
%\ea 
%
where $\nu'=\nu-\omega_L+\mu_3-\mu$, and $f(x)$ is the Fermi
distribution function. 
$A_3({\bf k},\nu)=2\pi \delta(\nu-(\epsilon_{\bf
  k}+\omega_{23}-\mu_3))$ and $A({\bf k},\nu) \equiv -2\,
\mathrm{Im}\, G({\bf k},\nu+i 0^+)$ are the spectral functions for
state $\ket3$ and $\ket2$, respectively.
%
%The spectral function for state $\ket3$ is
%$A_3({\bf k},\epsilon)=2\pi \delta(\epsilon-(\epsilon_{\bf
%  k}+\omega_{23}-\mu_3))$, and $A({\bf k},\epsilon) \equiv -2\,
%\mathrm{Im}\, G({\bf k},\epsilon+i 0^+)$ is the spectral function
%associated with state $\ket2$.  
%
Finally, one obtains
\ba
 I(\omega)&=&\frac{\bar{T}^2}{2\pi}\sum_{\bf k}
A({\bf k}+{\bf q}_L,\epsilon_{\bf k}-\omega-\mu)\nonumber\\
&&{} \times\left[f(\epsilon_{\bf k}-\omega-\mu)-f(\epsilon_{\bf
k}+\omega_{23}-\mu_3)\right] \,,
\label{eq:4}
\ea
where $\omega\equiv \omega_L - \omega_{23}$ is defined to be the RF detuning.

We have introduced a $T$-matrix formalism for addressing the effects
of finite temperature based on the standard mean field ground state
\cite{JS2a,ourreview}. Here the Green's function $G(\mathbf{k},\nu)$
contains two self-energy effects deriving from condensed Cooper pairs
as well as from finite momentum pairs, (which are related to pseudogap
effects).  These finite lifetime pairs have self energy $\Sigma_{pg}
({\bf k}, \nu) \approx \frac{\Delta_{pg}^2}{\nu +\epsilon_{\bf k} -\mu
  +i\gamma}$ where $\gamma \ne 0$. By contrast, the condensate which
depends on the superfluid order parameter (OP), $\Delta_{sc}$, enters
with $\Sigma_{sc}({\bf k}, \nu) = \frac{\Delta_{sc}^2}{\nu
  +\epsilon_{\bf k} - \mu }$. The resulting spectral function, which
can readily be computed from $\Sigma = \Sigma_{pg} + \Sigma_{sc}$, is
given by
\be 
A({\bf k},\nu)=\frac{2\Delta_{pg}^{2}\gamma
  (\nu+\xi_{\bf k})^2}{(\nu+\xi_{\bf
    k})^2(\nu^2-E_{\bf k}^{2})^2+\gamma^2(\nu^2-\xi_{\bf
    k}^2-\Delta_{sc}^{2})^2} \,.
\label{eq:5}
\ee
Here $\xi_{\bf k} = \epsilon_{\bf k}-\mu$.  $E_{\bf k} = \sqrt{
  \xi_{\bf k}^2 + \Delta^2 (T) }$ is the quasiparticle
dispersion, where $\Delta^2(T) = \Delta_{sc}^2(T) + \Delta_{pg}^2(T)$.
The precise value of $\gamma$, and even its $T$-dependence is not
particularly important, as long as it is non-zero at finite $T$. As is
consistent with the standard ground state constraints, $\Delta_{pg}$
vanishes at $T \equiv 0 $, where all pairs are condensed.  It is
reasonable to assume that $\gamma$ is a monotonically decreasing
function from above $T_c$ to $T=0$.  Above $T_c$, Eq.~(\ref{eq:5}) can
be used with $\Delta_{sc} = 0$.
Because the energy level difference $\omega_{23}$ ($ \approx 80 $ MHz)
is so large compared to other energy scales in the problem, the state
$\ket3$ is initially empty. It is reasonable to set $f(\epsilon_{\bf
  k}+\omega_{23}- \mu_3) = 0$ in Eq.~(\ref{eq:4}).

For the atomic gas in a trap, we assume a spherically symmetrical
harmonic oscillator potential $V(r) = m\bar{\omega}^2 r^2/2$ in our
calculations, where $\bar{\omega}$ is the trap frequency.  The density,
excitation gap and chemical potential will vary along the radius. These
quantities can be self-consistently determined using the local
density approximation (LDA). Here one replaces $\mu$ by a spatially
varying chemical potential $\mu(r)\equiv \mu-V(r)$. The same
substitution must be made for $\mu_3$ as well.  At each point, one
calculates the superfluid order parameter $\Delta_{sc}(r)$, the
pseudogap $\Delta_{pg}(r)$ and particle density $n(r)$ just as for a
locally homogeneous system; an integration over $r$ is performed to
enforce the total particle number constraint.  Equations (\ref{eq:4})
and (\ref{eq:5}) can then be used to compute the local current density
$I(r, \omega)$ and then to obtain the total net current $I(\omega) =
\int \mathrm{d}^3r\, I(r, \omega) $.

\begin{figure}
\centerline{\includegraphics[clip,width=3.2in]{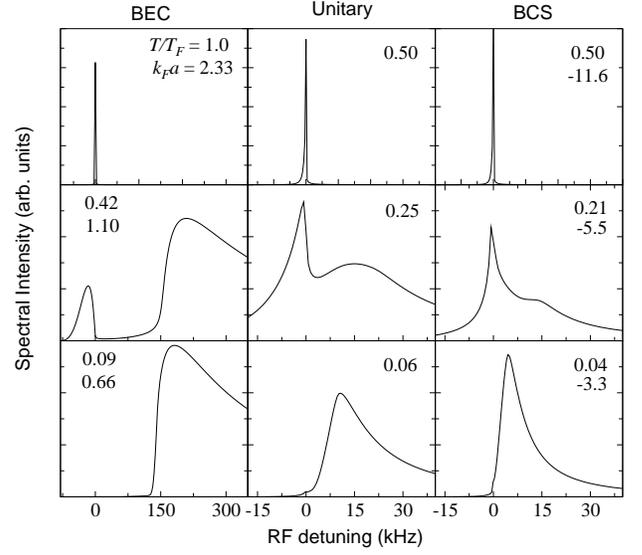}}
\caption{RF spectra $I(\omega)$ for the near-BEC (720 G, left column),
  unitary (837 G, $k_Fa = \infty$, middle column), and near-BCS (875
  G, right column) cases as a function of RF detuning $\omega$.  The
  values of $T$ (except in the first row) and $k_Fa$ were chosen to
  match the experimental values in Ref.~\cite{Grimm4a}.}
\label{fig:2}
\end{figure}

Figure \ref{fig:2} shows the calculated RF excitation spectra in a
harmonic trap for the near-BEC (720 G), unitary (837 G, $a=\infty$) and
near-BCS (875 G) cases, from left to right, for a range of $T$ from
above to below $T_c$.  Here, for definiteness, we take $\gamma/E_F = 0.1
+ 0.3T/T_c$, which monotonically increases with $T$ as one may expect
\cite{Chen4}.  The values of $T$ used in all rows but the first (which
involved the less interesting extreme Boltzmann regime) were chosen to
be consistent with the corresponding values of $T'$ used in
Ref.~\cite{Grimm4a}, on the basis of a theoretical thermometry
\cite{ChenThermo}.  Here $T'$ refers to the initial temperature of an
isentropic sweep starting from the BEC side of resonance. The values of
$k_Fa$ were calculated from the known values of $T_F$ and $a$.  Just as
in experiment \cite{Grimm4a}, two distinct maxima are seen. A very sharp
peak at $ \omega = 0$ appears only for $ T \neq 0$; this peak is, thus,
related to thermally excited fermion states.  A second and broader
maximum is present at sufficiently low $T$ and is connected to the
breaking of fermion pairs between states $\ket1$ and $\ket2$, with
subsequent transfer of state $\ket2$ to $\ket3$.  The broadening of zero
$\omega$ peak, as $T$ decreases, reflects the increasing values of the
gap $\Delta$. The general features of the spectra are in reasonable
agreement with experiment for all three cases shown.

The near-BEC plot is still far from the true BEC limit where $k_Fa$ is
arbitrarily small.  Nevertheless, one can see from the lowest $T$ figure
that the absorption onset is only slightly larger ($\sim 5 E_F$ as
compared to $\sim 4.6 E_F$) than the estimated two-body binding energy
$\hbar ^2/ma^2$, as expected.  This near-BEC figure makes it clear that
pairing effects are absent at the highest $T = 1.0 T_F\,\, (\gtrsim
T^*)$, where the free atom peak is symmetric and there is no sign of a
shoulder; this case is close to unitarity largely because of the size of
$k_F$.  It is also clear from the middle figure that the ``pairing gap"
forms above $T_c$, as is expected. Although not shown here, we find that
for the unitary case, there is an analogous pseudogap effect which
appears above $T_c$ via a shoulder in the spectra to the right of the
$\omega = 0$ peak. Only when $ T > T^*$ will this shoulder entirely
disappear. At unitarity, we find $T^* \approx 2 T_c$. Additionally, it
should be stressed that the near-BCS case is still very far from the
weak coupling BCS limit.

\begin{figure}
\centerline{\includegraphics[clip,width=2.5in,height=2.8in]{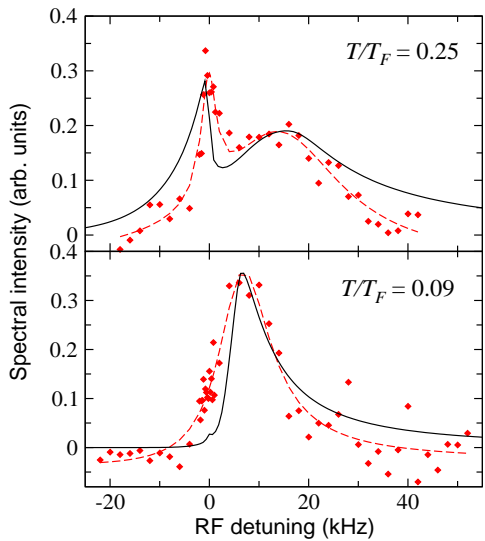}}
\caption{Comparison of calculated RF spectra (solid curve, $T_c
  \approx 0.29$) with experiment (symbols) in a harmonic trap
  calculated at 822 G for the two lower temperatures. The temperatures
  were chosen based on Ref.~\cite{Grimm4a}. The particle numbers were
  reduced by a factor of 2. }
\label{fig:1}
\end{figure}

Experimentally \cite{Grimm5a}, one defines the (averaged) ``pairing
gap'', $\Delta_{RF}$, as the energy splitting between the maximum in the
broad RF feature and the $\omega = 0$ point.  For the near-unitary case
in $^6$Li (at 822~G), $\Delta_{RF}/E_F \approx 0.25 $ at the
intermediate $T^\prime = 0.5T_F$, whereas at the lowest $T$ this ratio
is around $ 0.35$. The ratios found theoretically are roughly $0.35$ and
$0.38$ for these two cases.  However, when the field is increased to
precise unitarity (837~G) the numbers appear to considerably smaller
with a ratio of $\lesssim 0.2$.  On general grounds one can argue that
very little change is expected with these small changes in field near
unitarity.  Anharmonicity associated with a shallow Gaussian trap may
explain this small discrepancy, along with possible uncertainties in the
particle number \cite{private,Strinati5}.  There may also be some
interference with the Feshbach resonances between states $\ket1$ and
$\ket3$ and between $\ket2$ and $\ket3$ \cite{GrimmJuliennea}, which
overlap with the resonance between $\ket1$ and $\ket2$ but are not
included in the theory.

In Fig.~\ref{fig:1}, we compare our calculated spectra near unitarity
(solid curve) with experiment (symbols) at 822~G for the two lower
temperatures. The dashed curve is a fit to the data, serving as a
guide to the eye. As in Fig.~\ref{fig:2}, we calculated $k_Fa$ using
the experimental values of $T_F$ and $a$ \cite{Grimm5a}.  After
reducing the particle numbers by a factor of 2, as in
Ref.~\cite{Strinati5}, this brings the theory into very good agreement
with experiment.

\begin{figure}
\centerline{\includegraphics[clip,width=3.0in]{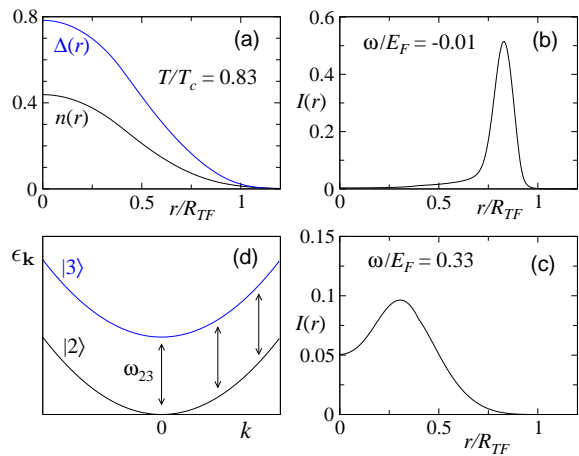}}
\caption{Origin of the two RF spectral peaks. Spatial distribution of
  (a) $\Delta$ and density $n(r)$, contributions to (b) the low
  frequency peak at $\omega/E_F=-0.01$ and (c) the high frequency peak
  at $\omega/E_F =0.33$ at unitarity in a harmonic trap, calculated at
  $T=0.23T_F$.  (d) Schematic transitions from state $\ket2$ to
  $\ket3$, showing the extended $k$-space contributions to the low
  energy peak, at the trap edge where $\Delta<T$.  Here $R_{TF}$ is
  the Thomas-Fermi radius for a noninteracting Fermi gas. }
\label{fig:3}
\end{figure}

To fully understand the RF excitation spectra, it is important to
determine where, within the trap, the two frequency peak features
originate.  Figure \ref{fig:3}(a) shows a plot of the density profile
$n(r)$ and excitation gap $\Delta(r)$ at unitarity and $T\approx 0.23T_F
\approx 0.83T_c$ as a function of radius.  Figures \ref{fig:3}(b) and
\ref{fig:3}(c) indicate the radial dependence of the local current
$I(r,\omega)$
%the integrand in the total current $I = \int I(r)\; d^3r$
for (b) frequency near zero, where there is a sharp peak, and for (c)
frequency near the pairing gap energy scale, where there is a broad
peak.  Just as conjectured in previous papers \cite{Grimm4a,Torma2}, it
can be seen that the low frequency peak is associated with atoms at the
edge of the trap.  These are essentially ``free'' atoms which have very
small excitation gap values, so that they are most readily excited
thermally.  By contrast, the pairing gap peak is associated with atoms
somewhere in the middle of the trap. 
% The radii corresponding to the two
%maxima of Figs.~\ref{fig:3}(b) and \ref{fig:3}(c) are indicated by
%arrows on the density profile of Fig.~\ref{fig:3}(a).

One might expect a rather broad free atom peak, reflecting a range of
values of $\Delta(r)$ at trap edge, but this peak is, in fact, quite
sharp, as is its experimental counterpart.  The sharpness of the free
atom peak is addressed via the schematic diagram of Fig.~\ref{fig:3}(d).
When $\Delta < T$ (as in the trap edge region), the dispersion of state
$\ket2$ reduces to a simple parabola as for free fermions; it is thus
similar to that of state $\ket3$, as seen in Fig.~\ref{fig:3}(d).
Momentum conservation leads to vertical transitions shown by the arrows
on the figure.  It is important to contrast this picture with the
situation for SN tunneling; here Pauli blocking effects are absent since
the final state is empty for all $\mathbf{k}$. As a result there is an
extended volume of $k$-space contributing to the transition at
$\omega_L=\omega_{23}$ (which corresponds to detuning $\omega = 0$),
thereby leading to the sharp spectral peak. At the very high $T$ which
were probed experimentally in Ref.~\cite{Grimm4a}, the sharp $\omega=0$
peak results in a similar fashion, although in this Boltzmann regime,
the gap $\Delta$ is completely irrelevant.

A plot analogous to Figs. \ref{fig:3}(b) and \ref{fig:3}(c) can be
made for the near-BEC case as well, to determine where in the trap the
RF gap $\Delta_{RF}$ arises.  It is easy to see that the threshold
region is associated with the trap edge where $\Delta(r)$ is small.
Indeed, when $\mu<0$, the excitation gap is given by $\sqrt{\mu^2
  +\Delta^2}$. This implies that the two-body binding energy $E_b
\approx -2\mu + a\pi n\hbar^2/m$ sets the scale for this threshold, in
much the same way as found for the deep BEC where $a\rightarrow 0$. As
long as $\mu < 0$, the values of $\Delta_{RF}$ and $E_b$ are very
close, becoming equal when $\Delta \rightarrow 0$ at the trap edge.
This supports the more detailed two-body analysis of these threshold
effects in the BEC presented in Ref.~\cite{ChengJulienne}.  However,
it should be stressed that there is an intrinsic rounding around the
threshold, as seen in the bottom left panel of Fig.~\ref{fig:2}.

At $T=0$ these LDA-based RF calculations can be compared with the
results \cite{Griffin5} of Bogoliubov-de Gennes (BdG) theory.  It
should be noted that BdG theory is appropriate for the particular mean
field ground state under consideration, but it cannot be applied at $T
\neq 0$, since it does not take into account the noncondensed bosonic
degrees of freedom.  A comparison presented in Ref.~\cite{Griffin5}
between a BdG calculation and its LDA approximation showed a
difference in the low frequency tunneling current at $T=0$ in the
fermionic regime ($\mu > 0$).  The finite spectral weight at precisely
$\omega=0$ in the BdG result was interpreted to arise from Andreev
bound states \cite{TormaNote}. 
%
%It should be noted that a small particle number $N$ and
%a narrow Feshbach resonance was used in the BdG calculations, and this
%situation may not be directly comparable to the present calculations
%even at $T=0$ \cite{TormaNote}.

It was also speculated that at finite $T$, Andreev effects may be
playing a role so that the free atom peak is possibly of a different
origin from that considered here and elsewhere \cite{Grimm4a,Torma2}.
It was noted in Ref.~\cite{Griffin5} that the BdG equations show that
the entire trapped gas is in the superfluid state below $T_c$, with
$\Delta_{sc}(r)$ being finite everywhere $n(r)$ is finite.  Therefore,
it was presumed that the free atom peak found in Ref.~\cite{Torma2} was
an artifact of the LDA at $T \neq 0$, since in this approximation, there
is a region of the trap where $\Delta_{sc} = 0$.

In support of the present viewpoint it is important to note that the
free atom peak derives from states where $\Delta(r) < T$.  The gap
$\Delta$, is the important energy scale, not the order parameter
$\Delta_{sc}$, for characterizing fermionic single-particle excitations.
This can be seen from the fact that the spectral function of
Eq.~(\ref{eq:5}) which enters into the RF calculations depends on
$\Delta$ through $E_{\bf k}$, and is not particularly sensitive to
$\Delta_{sc}$.  From Fig.~\ref{fig:3}(a) it follows that $\Delta$ is
finite wherever $n(r) \ne 0$.  It behaves similarly to the nonvanishing
order parameter in BdG-based calculations.  Furthermore, both the OP in
Ref.~\cite{Griffin5} and $\Delta$ (for the present case) behave as
$\Delta \sim \exp(-\pi/2k_F|a|)$, becoming exponentially small at the
trap edge.
Thus, as a result of pseudogap effects (which serve to distinguish
$\Delta$ and $\Delta_{sc}$ at any finite $T$) we believe that the
concerns raised earlier \cite{Griffin5} about the applicability of LDA
for addressing RF experiments are not warranted.

The results of this paper support a previous theoretical interpretation
\cite{Torma2} of RF experiments \cite{Grimm4a} in the 
%intermediate or
unitary regime, which applied the pseudogap based formalism of the
present paper, albeit with approximated spatial density and gap
profiles. The present calculations avoid these approximations, and
lead to spatial density profiles \cite{JS5} and related thermodynamics
\cite{ThermoScience} which are in good quantitative agreement with
experiment \cite{JS5}.  Our work clarifies the origin of the two
generic peak structures seen in RF experiments, and addresses the
entire magnetic field range which has been studied experimentally.
The zero frequency peak comes from atoms at the edge of the trap,
where $\Delta < T$.  We interpret these as ``small-gap" rather than
``in-gap" excitations. We have shown that the sharpness of this peak
is associated with an extended momentum space available for the
$\omega=0$ excitations.  The broader peak derives from the breaking of
pairs and, except in the extreme BCS limit, this peak is present above
$T_c$, reflecting pseudogap effects.

We are extremely grateful to Cheng Chin, Rudi Grimm and Paivi T\"orm\"a for
many helpful discussions.  This work was supported by NSF-MRSEC Grant
No.~DMR-0213745 and by the Institute for Theoretical Sciences
(University of Notre Dame and Argonne Nat'l Lab), and by DOE, No.
W-31-109-ENG-38 (QC).

%\bibliographystyle{apsrev}
%\bibliography{Review2}

\begin{thebibliography}{25}
\bibitem[{Tor()}]{TormaNote}
\bibinfo{note}{T\"orm\"a has suggested that the BdG calculations based on a
  small particle number $N$ and a narrow Feshbach resonance may not be directly
  comparable to the LDA calculations even at $T=0$.}
\end{thebibliography}

\end{document}